\documentstyle[pra,aps,twocolumn,epsf]{revtex}
\begin{document}
%\narrowtext
\draft

\title{Spin domain formation in spinor Bose-Einstein condensation}
\author{
  Tomoya Isoshima,\thanks{ Electronic address: tomoya@mp.okayama-u.ac.jp }
  Kazushige Machida, and
  Tetsuo Ohmi\ \footnotemark[2]
}
\address{
  Department of Physics, Okayama University, Okayama 700-8530, Japan\\
  \footnotemark[2]Department of Physics, Graduate School of Science,
  Kyoto University, Kyoto 606-8502, Japan
}
\date{\today}
\maketitle

\begin{abstract}
The spatial structure of the spinor Bose-Einstein condensates
with the spin degrees of freedom is analyzed
based on the generalized Gross-Pitaevskii equation (GP)
in the light of the present spin domain experiment
on $m_F=\pm 1$, and  $0$ of the hyperfine state $F=1$ of $^{23}$Na atom gases.
The GP solutions in three- and one-spatial dimensional cases reproduce
the observed spin domain structures,
revealing the length scale associated with the existence of
the weak interaction of the spin-spin channel,
other than the ordinary coherence length related to the density-density channel.
The obtained domain structure in GP is compared
with the result in Thomas-Fermi approximation.
The former solution is found to better describe
the observed features than the latter.
\end{abstract}


%==================================================================
\section{Introduction}

The experiments for Bose-Einstein condensation (BEC) in alkali atomic gases,
such as $^{87}$Rb\cite{cornell}, $^{23}$Na\cite{ketterle},
and $^{7}$Li\cite{hulet},
have been
performed under a strong magnetic field
since its experimental realization in alkali atomic gases in 1995.
They used the magnetic field to confine a BEC system.
Because the atom spin direction adiabatically follows the magnetic field,
the spin degrees of freedom are frozen in these magnetic
trapping experiments\cite{DalfovoReview}.

Recently, the MIT group has succeeded in creating BEC
by an optical dipole trap formed by a single
infrared laser beam\cite{stumper,StengerNature,miesner}
in which the spin degrees of freedom are
all active ($m_{F}=1,0,-1$ of the $F=1$ atomic hyperfine state for $^{23}$Na).
These atoms with the three hyperfine substates simultaneously undergo
Bose condensation, leading to a spinor BEC,
a situation analogous to
superfluid $^3\text{He}$\cite{vollhardt} of a neutral Fermion system
or a triplet superconductor of a charged Fermion system
such as UPt$_{3}$\cite{machida}.
The spin degrees of freedom play a fundamentally important role
for governing their physics.
An advantage of the present spinor BEC over $^3\text{He}$ or UPt$_{3}$
is that it is a weakly interacting system and
we know
%--------------may11
fairly
%--------------may11
well how the (quasi) particles interact
%
%--------------may11
using the knowledge of atomic physics.
% because there are no crystal lattice nor the Cooper pairs.
%--------------may11
%
These facts allow us to have a chance
to construct a microscopic many-body theory from first principles,
using only a few fundamental matter parameters.

Stimulated by the earliest optical trap experiment
by Stumper-Kurn {\it et al.}\cite{stumper},
general theoretical frameworks for describing a spinor BEC
were given independently
by Ohmi and Machida\cite{OhmiMachida} and by Ho\cite{HoSpinor}.
They are equivalent basically.
The framework is based on Bogoliubov theory which is extended to
a vectorial order parameter with three components,
corresponding to $m_{F}=1,0,-1$ of the $F=1$ atomic hyperfine state,
giving rise to generalized Gross-Pitaevskii (GP) equation.
They calculate low-lying collective modes
such as sound wave, spin wave, and their coupled mode
and predict various topological defect structures, or spin textures.
We also note that there are many remarkable researches\cite{phaseseparation}
with the phase separation problem of a two-component BEC
consisting of different hyperfine states.

Subsequently, Stenger {\it et al.}\cite{StengerNature} produce
%-----------------
% a spinor BEC with a long cigar shape
% whose aspect ratio is over 40
% trapped optically
% %
an optically trapped spinor BEC with a long cigar shape
whose aspect ratio is over 40
%-----------------
and examine whether these three component BEC
can be either miscible or immiscible.
The external magnetic field is applied along the long axis (the $z$ axis)
to see the spin domain  formation.
After releasing the spinor BEC,
the Stern-Gerlach separation of the cloud is performed
to reproduce the original domain structure.
Their analyses are based on the Thomas-Fermi (TF) approximation,
which neglects the kinetic energy,
of the above generalized GP equations.
They conclude that
the spin-dependent interaction channel of the present spinor BEC
is antiferromagnetic,
rather than ferromagnetic in this particular hyperfine state.

In this paper,
we analyze the above experiment
by Stenger {\it et al.}\cite{StengerNature} in more detail
to determine the three-dimensional structure of the domain wall 
including the miscible and immiscible spin domain structures.
%The actual calculations are done both for spatially three-dimensional
The one-dimensional calculation is also done to discuss
%
%------------------
the characteristic lengths.
% the coherence lengths.
%------------------
%
The arrangement of the paper is as follows:
In next section, we introduce the generalized Gross-Pitaevskii  equation of a Bose-Einstein 
condensed system with internal spin degrees of freedom.
In Sec.\ \ref{sec:threeDim} we simulate the actual experimental system
in light of the experimental conditions of Ref.\ \cite{StengerNature}
and investigate three-dimensional systems.
The properties of a one-dimensional system
in an idealized situation are explored in Sec.\ \ref{sec:oneDim}.
The last section is devoted to summary and conclusion.

%==================================================================
\section{Formulation}

The Hamiltonian invariant under spin space rotation and gauge transformation is written in terms of the three component field operators: $\Psi_{+1}, \Psi_{0}, \Psi_{-1}$, corresponding to  the sublevels $m_F=+1, 0, -1$ of the hyperfine state $F=1$.
Namely, it is given by 
\begin{eqnarray}
  {\rm H}
  &=&
  \int \!\!  d{\bf r} \sum_i
  \Psi^{\dagger}_i({\bf r}) h_i({\bf r}) \Psi_i({\bf r})       \nonumber
\\&&
  +\frac{g_n}{2} \sum_{i,j}
  \Psi_i^{\dagger}({\bf r})
  \Psi_j^{\dagger}({\bf r})
  \Psi_j({\bf r})
  \Psi_i({\bf r})                   \nonumber
\\&&
  +\frac{g_s}{2} \sum_{\alpha}
  \Biggl(
    \sum_{i,j}
    \Psi_i^{\dagger}({\bf r})
    (F_{\alpha})_{i,j}
    \Psi_j({\bf r})
  \Biggr)^2,     \label{eq:Hamil}
\end{eqnarray}
where 
\begin{equation}
  h_i({\bf r})=
    -\frac{\hbar^2\nabla^2}{2m} -\mu
    + V({\bf r}) - p(F_{z})_{i,i} + q(F_z^2)_{i,i}    \label{eq:HamilOne}
\end{equation}
is the one-body Hamiltonian,
$V({\bf r})$ is the trapping potential,
$i,j=0,\pm 1$ are spin indices and $F_{\alpha} (\alpha=x,y,z)$ are the following
$3 \times 3$ spin matrices:
\begin{eqnarray}
F_{x} &=& \frac{2}{\sqrt{2}} \left(
   \begin{array}{ccc}
      0 & 1 & 0 \\
      1 & 0 & 1 \\
      0 & 1 & 0 
   \end{array}
\right),
\\
F_{y} &=& \frac{2i}{\sqrt{2}} \left(
   \begin{array}{ccc}
      0 & -1 &  0 \\
      1 &  0 & -1 \\
      0 &  1 &  0 
   \end{array}
\right),
\\
F_{z} &=&  \left(
   \begin{array}{ccc}
      1 & 0 &  0 \\
      0 & 0 &  0 \\
      0 & 0 & -1 
   \end{array}
\right).
\end{eqnarray}
We have introduced the linear and quadratic Zeeman energies\cite{StengerNature},
$E_{\text{ze}} = E_0 - \tilde{p}\langle F_z\rangle + q\langle F_z^2\rangle$
with $p \equiv \tilde{p} + p_0$ and
the Lagrange multiplier $p_0$  represents the conversation
of the total spin of the system.

The interaction constants are related to the two kinds of
the scattering lengths
 $a_0$ and $a_2$
%
%---------------may11
corresponding to the total spin zero channel and two channel:
%---------------may11
%
%
%
\begin{eqnarray}
  g_n &=& \frac{4 \pi \hbar^2}{m} \cdot \frac{a_0 + 2a_2}{3},
  \label{eq:g0}
\\
  g_s &=& \frac{4 \pi \hbar^2}{m} \cdot \frac{a_2 - a_0}{3}.   \label{eq:g2}
\end{eqnarray}
Let us introduce the three component order parameters:
$\phi_i=\langle \Psi_i\rangle$ with $i=x,y,z$.
Following the standard procedure \cite{FetterWalecka}, we can derive the 
Gross-Pitaevskii equation for describing $\phi_i({\bf r})$
from the Hamiltonian Eq.\ (\ref{eq:Hamil}): 
\begin{eqnarray}
   \left[
     h_i({\bf r}) + g_n \sum_{j} |\phi_j({\bf r})|^2
   \right]
   \phi_i({\bf r})    \qquad         \nonumber
&&\\
  + g_s \sum_{\alpha} \sum_{j,k,l}
    (F_{\alpha})_{i,j} \phi_j \phi_k^{\ast} (F_{\alpha})_{k,l} \phi_l
    &=& 0.               \label{eq:gp}
\end{eqnarray}
The energy of the system is given by
\begin{equation}
  E
  =
  \int \!\!  d{\bf r} \sum_i\left\{
    \phi^{\ast}_i({\bf r}) h_i({\bf r}) \phi_i({\bf r})
  \right\}
  + \int \!\!  d{\bf r}\{E_{n}({\bf r}) + E_{s}({\bf r})\}.          \label{eq:Ene}
\end{equation}
where
\begin{eqnarray}
  E_{n}({\bf r}) &=&
  \frac{g_n}{2} \sum_{i} |\phi_i({\bf r})|^4 ,       \label{eq:EneGn}   
\\
  E_{s}({\bf r}) &=&
  \frac{g_s}{2} \sum_{\alpha}
  \Biggl(
    \sum_{i,j}
    \phi_i^{\ast}({\bf r})
    (F_{\alpha})_{i,j}
    \phi_j({\bf r})
  \Biggr)^2.     \label{eq:EneGs}
\end{eqnarray}
We will use the following notation in Sec.\ \ref{sec:oneDim}:
$\zeta_{i}({\bf r}) = \frac{\phi_{i}({\bf r})}{\sqrt{n({\bf r})}}$ 
where $n({\bf r}) = \sum_i |\phi_{i}|^2$.

%==================================================================
\section{Three dimensional case}\label{sec:threeDim}

In this section we analyze the three-dimensional systems
in comparison with the actual experiments done by Stenger {\it et al.}\cite{StengerNature}.
The main interests are (1) the $z$-position of the domain wall,
and (2) the radial shape including the mutual overlapping of the spin domains. 

%------------------------------------------------------------------
\subsection{Experiment}
In order to establish a model system we briefly explain their
experimental conditions.
The condensate is under the magnetic field $B(z)$ which is applied along the $z$ axis.
Since the atoms are in the hyperfine $F=1$ state, the  condensate consists of the three 
components: $\phi_{+1}, \phi_{0}$, and $\phi_{-1}$.
The non-uniform magnetic field $B(z)$ is characterized by its field gradient: $p^\prime \propto dB(z)/dz$.
The $p$ in $E_{\text{ze}}$ is set to zero at $z=0$ and $p=p^{\prime} z$.
The stronger-field side of the cigar-shaped condensate is filled with the $+1$ 
condensate.
As the system follows the total spin conversation, the opposite side is 
filled with the $-1$ condensate.
Stenger {\it et al.}\cite{StengerNature} observe the column density distribution of the
three spin components and estimate the interaction constant $g_s$ from the $z$-position of the 
domain wall between the spin components.
The result depends not only on $p$,  but also on the coefficient $q$.
$q\propto B_0^2$ where $B_0 \simeq B(z)$ is the base magnetic field.
The experiments are done for various $p$'s and $q$'s.

They estimate the relationship of $p$ and $q$ with the $z$ position of the 
domain wall by approximating the cigar-shaped system as a one-dimensional 
one with uniform density, which is assumed to be $2/3$ times the peak density.
The kinetic term is also ignored.
This allows us to draw lines in $p\text{-}q$ 
phase diagram as a function of the interaction constant $g_s$ (in their paper, $c_2$).

%------------------------------------------------------------------
\subsection{Calculation}
In order to check whether or not their analysis based on the above assumptions
(1-dimensional TF approximation, 1dTF)
agrees with the full 3-dimensional calculations of GP equation,
%(i.e.\ including the kinetic term),
we determine the spatial profiles of the 
rotationally-symmetric (around the $z$ axis)
spinor condensate by solving the GP equation in the $r\text{-}z$ space using
the difference between the scattering lengths
$a_2 - a_0 = 0.19 \text{nm}$ ( $\propto g_s$ )
%
%
% and the peak density $4.35 \times 10^{20} m^{-3}$
%
and the peak density $n_{\text{peak}} = 4.35 \times 10^{20} m^{-3}$
they estimated.
Assuming this,
we compare the following three calculations by 1dTF,
by 3-dimensional TF approximation (3dTF), and
by 3-dimensional GP equation (3dGP).
In 1dTF, it is assumed that the density is
$\frac{2}{3} n_{\text{peak}}$.

\begin{figure}
  \epsfxsize=8cm \epsfbox{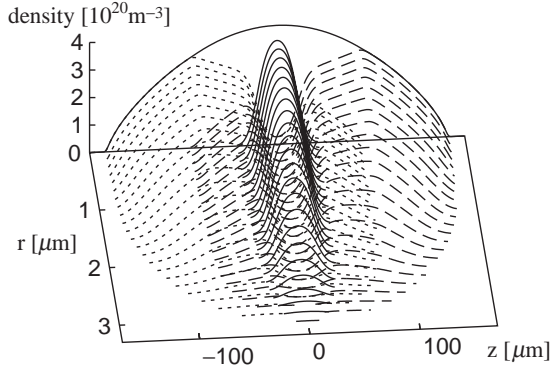}

%-----------------------------
  \caption{
    The density plot of the $r\text{-}z$ plane.
    The dashed lines, the solid lines around $z=0$, and the dotted line
    mean the densities of the $+1$, $0$, and $-1$ components.
    The upper solid curve at $r=0$ indicates the total density at $r=0$.
    The peak density is $4.35\times 10^{20} m^{-3}$,
    $p^{\prime}=1.0 \text{Hz}/\mu m $, and $q = 2.0 \text{Hz}$.
    The optical confinement is assumed as the harmonic potential given by
%
%   $V({\bf r})={1\over 2}(\nu_r r^2+\nu_z z^2)$,
  $V({\bf r}) =
  \frac{m}{2}\{ (2\pi\nu_r r)^2 +(2\pi\nu_z z)^2 \}$,
    where $\nu_z=15\mbox{Hz}$ and $\nu_r=900\mbox{Hz}$.
    The total particle number is $8.71 \times 10^5$.
  \label{fig:3dGPdns}
  }
\end{figure}

\begin{figure}

  \epsfxsize=8cm \epsfbox{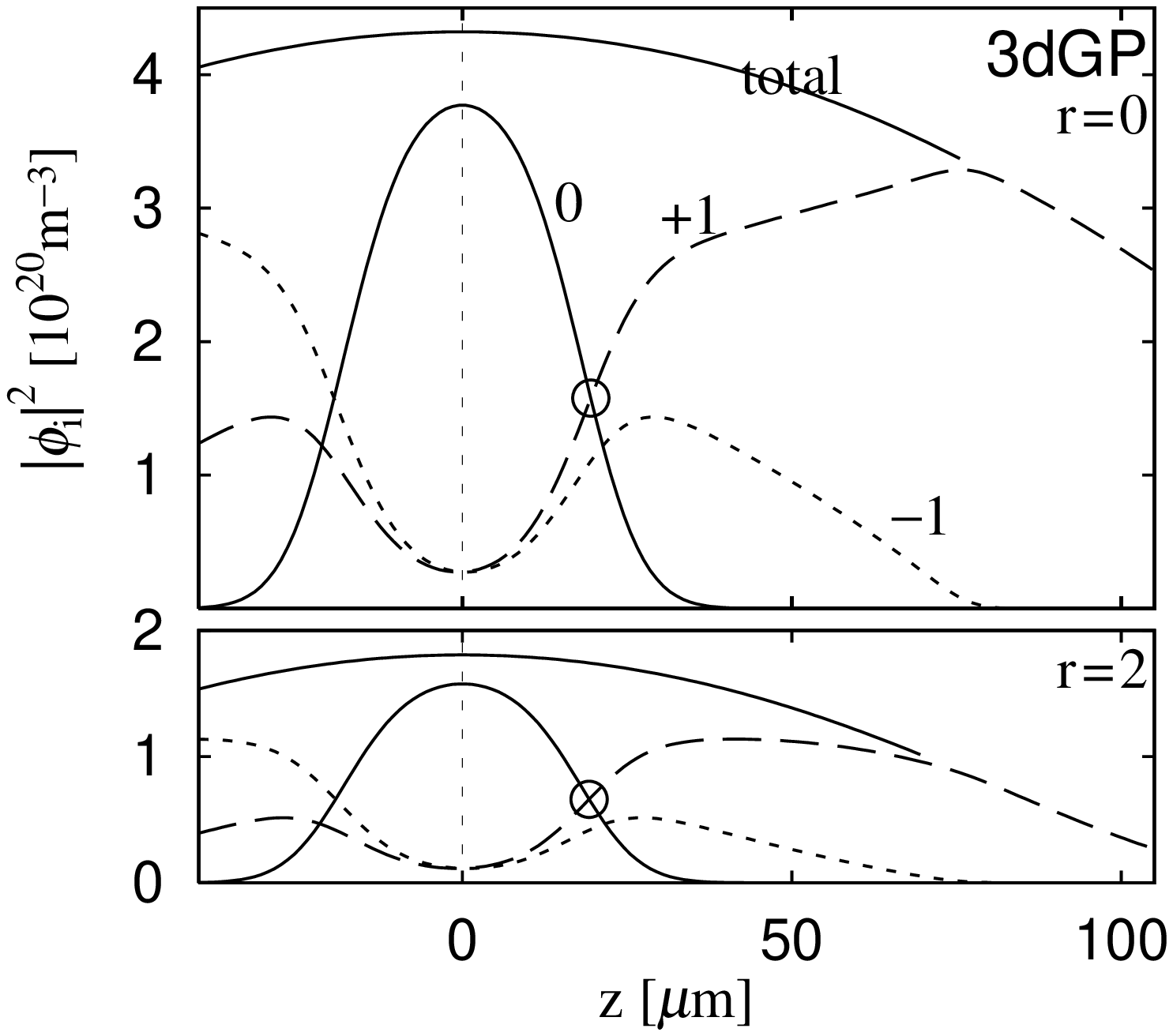}

  (a)

  \epsfxsize=8cm \epsfbox{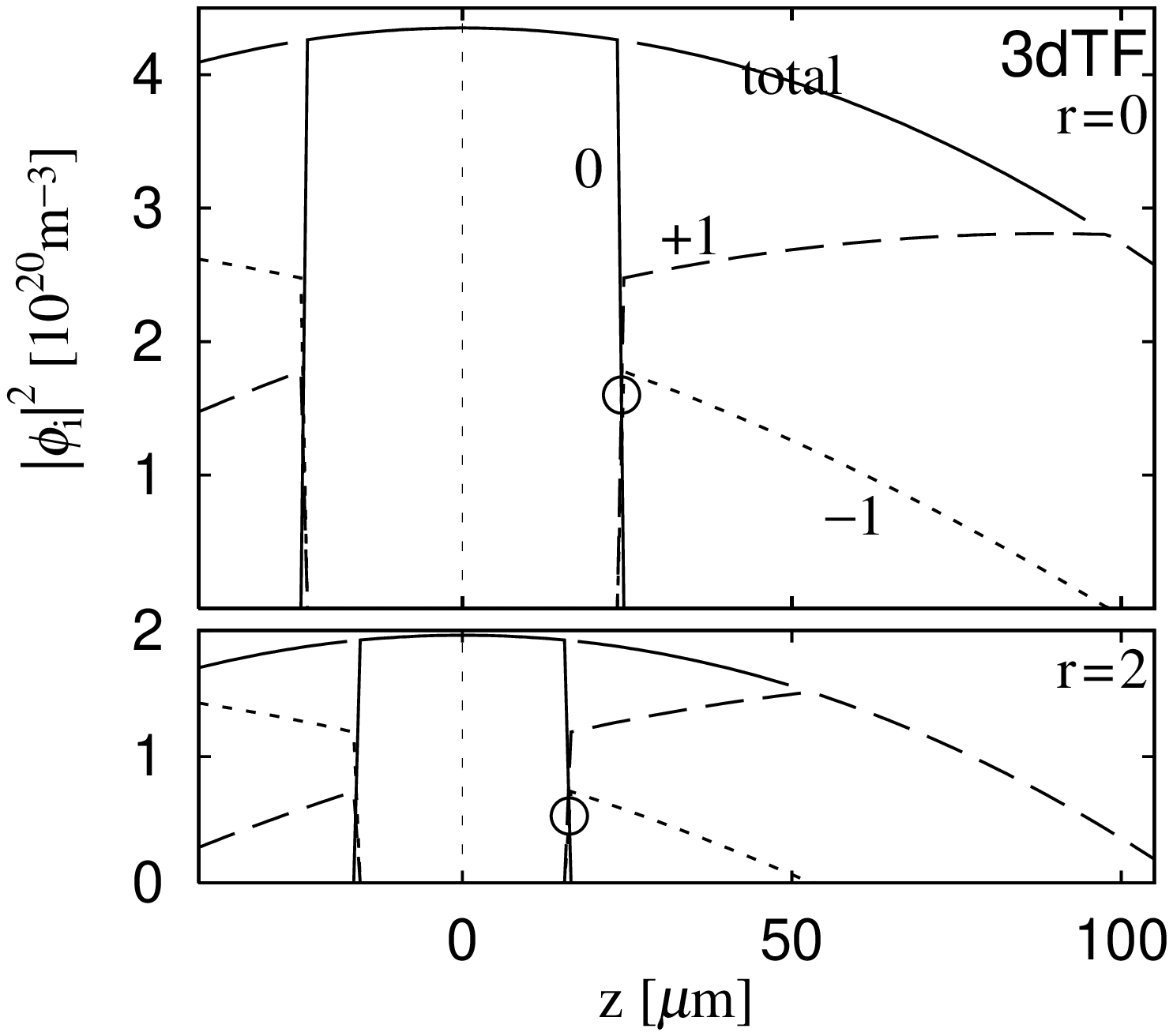}

  (b)
%-----------------------------
  \caption{
    The density profiles of the condensate.
%%  The peak density, $p$, $q$, and the shape of the harmonic
%%  confinement are equivalent to Fig.\ \protect\ref{fig:3dGPdns}
%%  [Only the total particle number in (b),
%%  which is $8.99 \times 10^5$, is different].
The peak density, $p$, $q$, and the shape of the harmonic
confinement are same as in Fig.\ \protect\ref{fig:3dGPdns}
except for the total particle number $8.99 \times 10^5$ in (b).
    The dashed lines, the solid lines around $z=0$, and the dotted line
    mean the densities of the $+1$, $0$, and $-1$ components.
    The upper solid curves indicate the total density.
    (a) The density profile at $r=0 \text{\ and\ } 2\mu m$. 
    The $\pm 1$ components exist even at $z=0$.
    We call the crossing between the $0$ and $+1$ components (circles)
    the domain wall.
    The $z$ coordinates of the domain wall are 
    $z = 19.5 \mu m$ at $r=0$ and $z = 19.2 \mu m$ at $r=2\mu m$.
    (b) The density profile when calculated with 3dTF.
    As $r$ increases, the domain wall (circles) between the
    $0$ component and $\pm 1$ components moves toward $z=0$.
    The $z$ coordinates of the domain wall are 
    $z = 24.1 \mu m$ at $r=0$ and $z = 16.2 \mu m$ at $r=2\mu m$.
  \label{fig:densGP}
  }
\end{figure}

We first discuss the domain structure along the $r$ direction
and the mutual overlapping between the three components.
One example ($p^{\prime}=1.0 \mbox{Hz}/\mu m $, $q=2.0\mbox{Hz}$) of the calculated results with 3dGP 
is depicted in Figs.\ \ref{fig:3dGPdns} and \ref{fig:densGP}(a).
There is a large 
overlapping region consisting of the +1 and the -1 components,
between the region consisting of the +1 (or -1) component only and of the 0 component only.
In other words,
the double peak structures of the $-1$ component and $+1$ component are
situated both at the $z>0$ side and $z<0$ side
in Figs.\ \ref{fig:3dGPdns} and \ref{fig:densGP}(a).
This feature accords with the observation (Fig.\ 3 in \cite{StengerNature}).
Fig.\ \ref{fig:densGP}(a)
also shows that the condensate itself
allows the overlapping of 3 components at $z=0$
in certain values of $p^{\prime}$ and $q$.
%
%%\ \cite{overlap}.
%%%%---------move to Ref.
% \bibitem{overlap}.
% although this kind of overlapping observed experimentally
% may mainly arise from the processes of observing
% or releasing from the optical trap.
%%%%---------move to Ref.
%
This existence of three components at $z=0$ does not
occur in calculations with 3dTF [e.g.\ Fig.\ \ref{fig:densGP}(b)].
Usually the difference between the results of TF approximation and of full 
GP equation is considered to be an order of the coherence length $\sim1\mu m$ here, but the 
difference between the two treatments is as long as $50 \mu m$, which is far longer than expected.
This behavior also occurs in the one-dimensional system and is discussed 
thoroughly at Sec.\ \ref{sec:oneDim}.
(This overlapping at $r=0$ does not occur at larger $q$, but the 
length scale stays as long as $10 \mu \text{m}$.)

\begin{figure}
  \epsfxsize=8cm \epsfbox{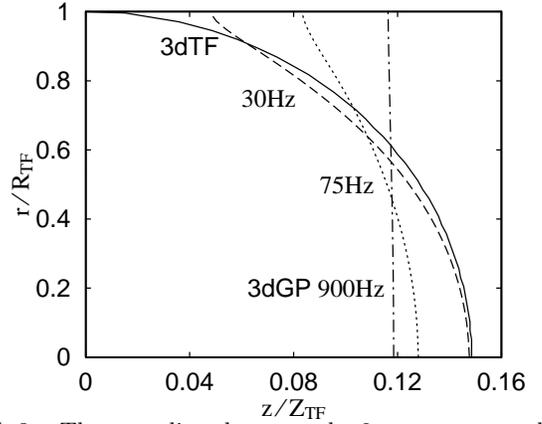}
  \caption{
    The cross-lines between the $0$ component and the $+1$ component
    for $\nu_r = 30, 75$, and $900 \text{Hz}$.
    The solid line, which is independent of $\nu_r$, is the line when
    TF approximation is used.
    The $900 \text{Hz}$ line correspond to
    Fig.\ \protect\ref{fig:3dGPdns}.
    Each axis is normalized by the TF radius and length,
    which are defined as
$
%%%%%%%  $4.35 \times 10^{20} m^{-3} \times g_n$.
  \text{(peak density)}\times g_n
  = \frac{m}{2} (2 \pi \nu_r R_{\text{TF}} )^2 
  = \frac{m}{2} (2 \pi \nu_z Z_{\text{TF}} )^2 
$.
    The $R_{\text{TF}}$ is $82.2, 32.9, \text{\ and\ } 2.74 \mu m$
    for $\nu_r = 30, 75, \text{\ and\ }  900 \text{Hz}$ respectively.
 $Z_{\text{TF}}$ is $164 \mu m$.
  \label{fig:cross}
  }
\end{figure}

Another interesting difference between 
the results from 3dTF and 3dGP exists at the crossing 
point between the 0 component and the +1 component.
As seen in Fig.\ \ref{fig:densGP}(a), the $z$ coordinates of the crossing points do not 
move appreciably between the two cases: $r=0$ and $r=2$.
However, the crossing point moves significantly in TF as seen from Fig.\ \ref{fig:densGP}(b).
The trace of these crossing points in the $r\text{-}z$ plane
for various radial harmonic potentials are depicted in Fig.\ \ref{fig:cross}.

%------------------------------------------------------------------
\subsection{Phase diagram}
\begin{figure}
\widetext
  \epsfxsize=12cm \epsfbox{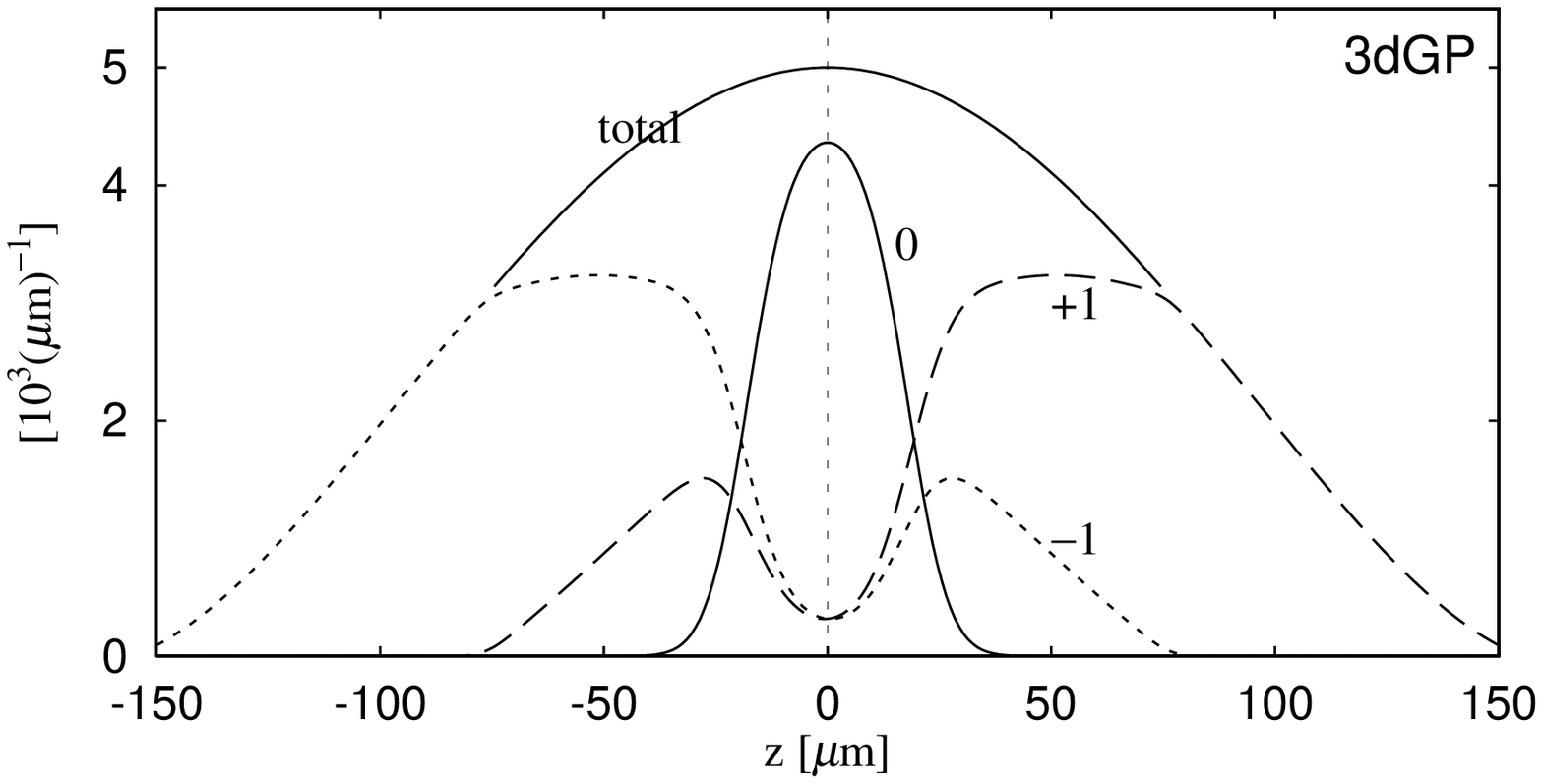}

  (a)

  \epsfxsize=12cm \epsfbox{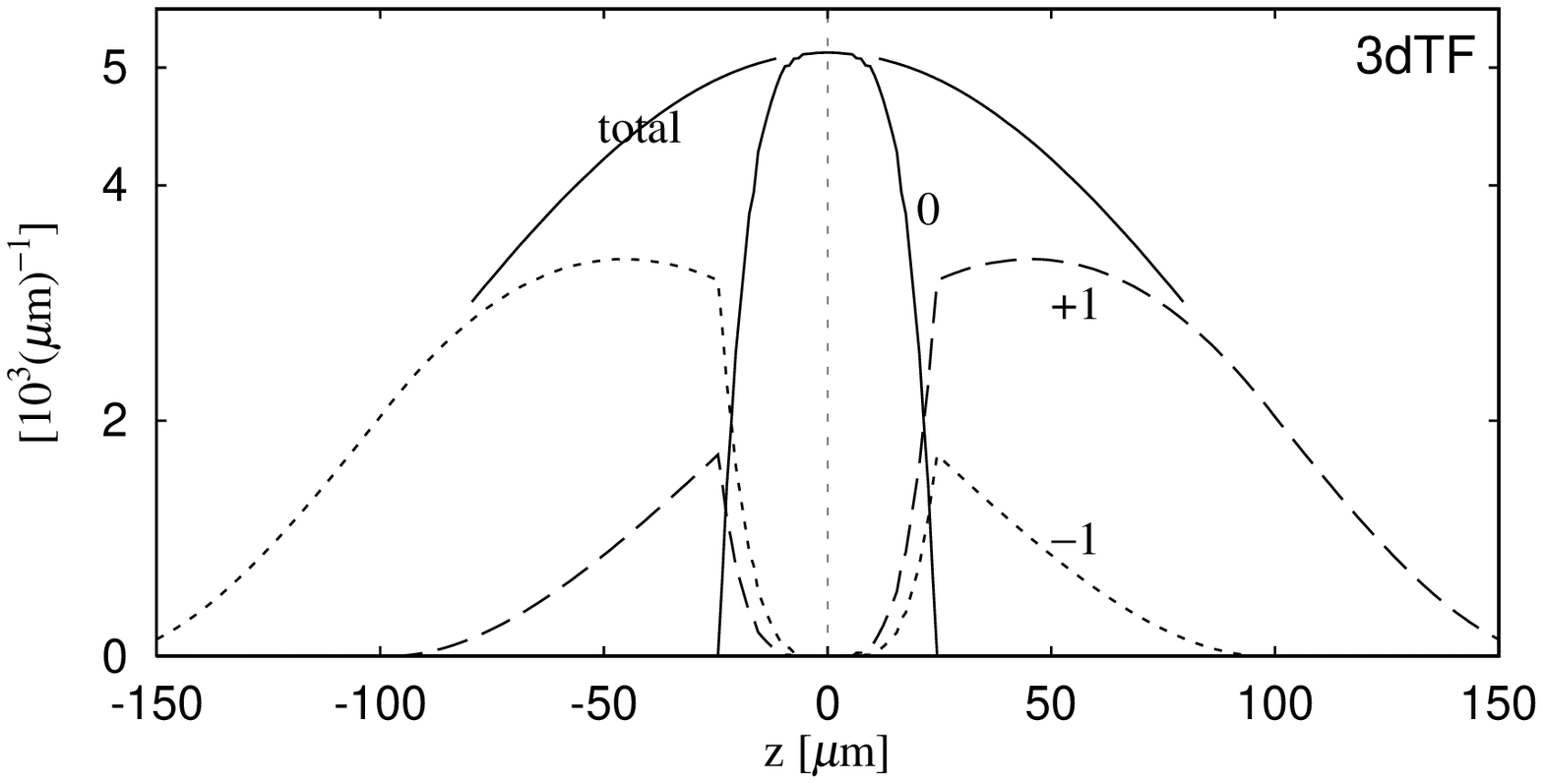}

  (b)
  \caption{
    (a) The $r$-integrated density profiles for
    $p^{\prime}=1.0 \text{Hz}/\mu m $ and $q = 2.0 \text{Hz}$.
    The overlapping of these integrated densities
    simply reflect the shape of Fig.\ \protect\ref{fig:densGP} (a).
The $z$ coordinate of the domain wall is $19.3\mu m$.
    (b) The corresponding result for 3dTF.
    The overlapping of these integrated density
    comes from the $r$-dependent shift of the domain wall
    depicted in Fig.\ \protect\ref{fig:cross}.
The $z$ coordinate of the domain wall is $21.5\mu m$.
    These figures should be compared with Fig.\ 3
    in \protect\cite{StengerNature}.
  \label{fig:integ}
  }
\narrowtext
\end{figure}

To compare the $z$-position of the domain wall between the 0 component and the 1 component,
we integrate the density profile e.g.\ Fig.\ \ref{fig:3dGPdns} along the $r$ surface.
Figure\ \ref{fig:integ} (a) shows an example of the results in 3dGP
and Fig.\ \ref{fig:integ} (b) shows that in 3dTF.
When the density profile is integrated along the $r$ direction
this produces further overlapping between the components
especially in 3dTF
[compare Fig.\ \ref{fig:densGP}(b) with Fig.\ \ref{fig:integ}(b)].
As seen from Fig.\ \ref{fig:integ} (a),
the $z$ coordinate of the crossing point in 3dGP
is slightly smaller than the 3dTF case in Fig.\ \ref{fig:integ} (b).

\begin{figure}
  \epsfxsize=8cm \epsfbox{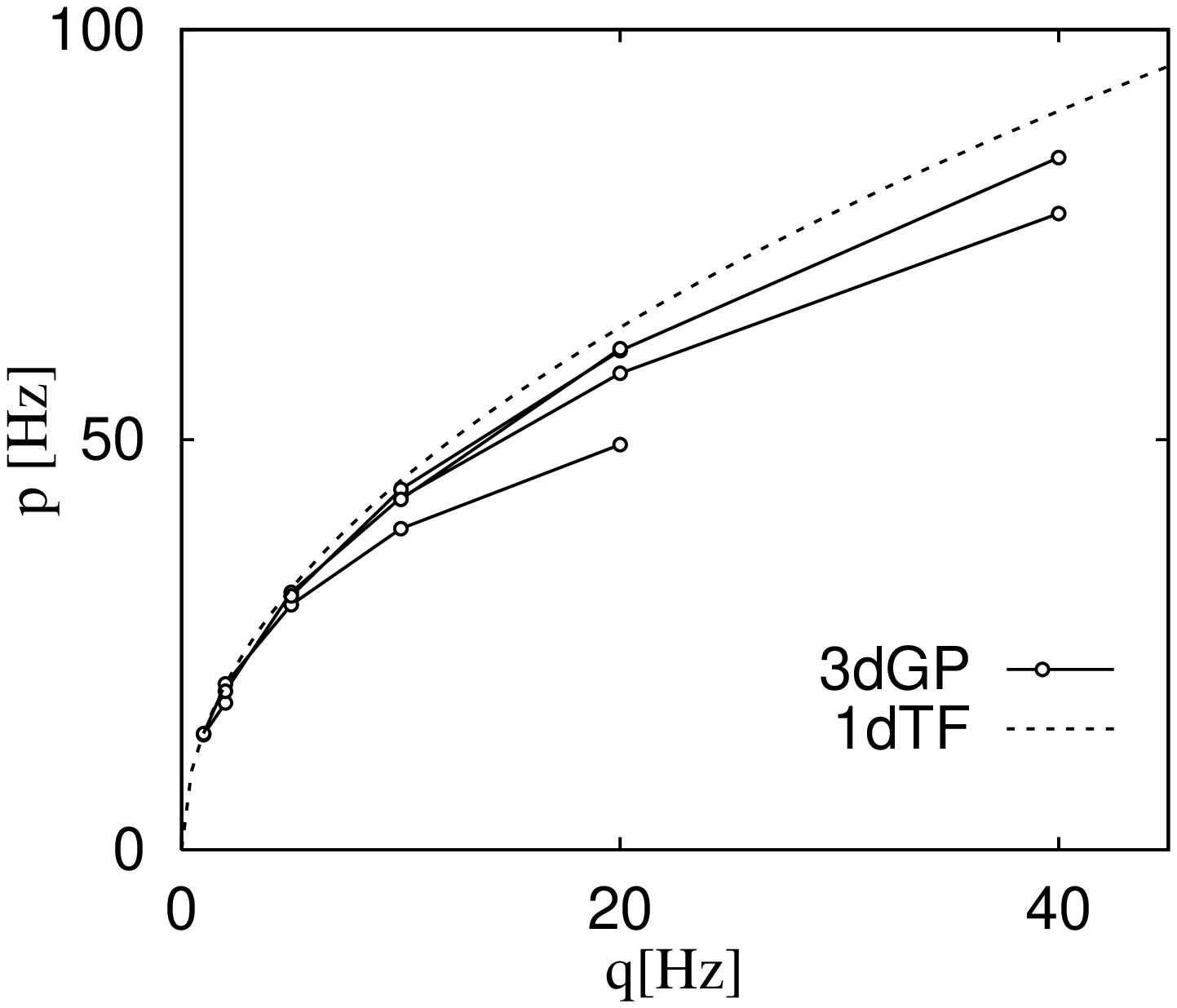}
 
  (a)

  \epsfxsize=8cm \epsfbox{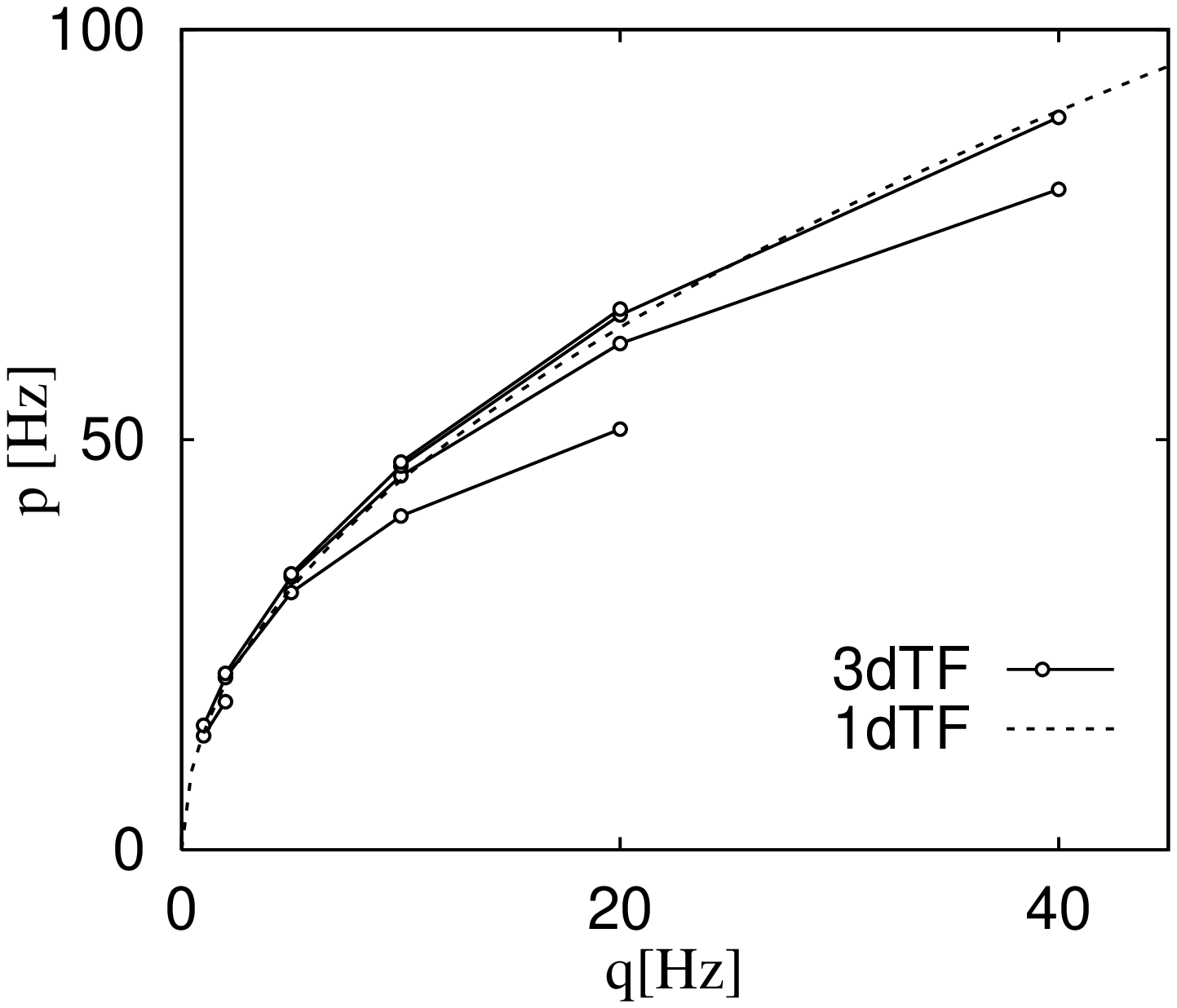}

  (b)
  \caption{
    The $p\text{-}q$ diagram.
    (a) The comparison between the the results in 1dTF and 3dGP.
    (b) The comparison between the results in 1dTF and 3dTF.
    The magnetic field parameters are
    $p^{\prime} = 0.2, 0.5, 1, 2$, and $4 \text{Hz}/\mu m$, and 
    $q = 1, 2, 5, 10, 20$, and $40 \text{Hz}$
    (not all of the combinations are used).
    The 1dTF line is $p = 2\sqrt{50.7q}$.
  \label{fig:cpq}
  }
\end{figure}

The acquired $z$ value of 3dTF
for various $p^{\prime}$ and $q$ values are shown in Fig.\ \ref{fig:cpq}(a).
The 1dTF line $p = 2\sqrt{50.7q}$ is also plotted.
The value ``50.7'' is derived uniquely from the assumed scattering
lengths and the peak density ($\times 2/3$).
No significant difference is seen.
The lower branch lines in Fig.\ \ref{fig:cpq} comes from
the density profile along the $z$ axis;
When $z$ is large, the value ``50.7'' ($\propto$ density)
must become much smaller.
Figure\ \ref{fig:cpq}(b) compares the 1dTF and the 3dTF.

The differences in $p\text{-}q$ diagram between 1dTF, 3dTF, and 3dGP are small.
Therefore we conclude that 
the estimate in Ref.\ \cite{StengerNature} is correct.
But the shapes of domain walls are not simple  
as shown in Figs.\ \ref{fig:densGP} and \ref{fig:cross}.

%==================================================================
\section{One Dimensional case}\label{sec:oneDim}

We have seen in the previous section that the difference between
the density distributions by the TF result and GP result 
is much wider than the usual coherence length $\xi \lesssim 1 \mu m$.
In this section, we take up the one-dimensional system to investigate
why this is so, taking into account
the effects of the kinetic term
and the interaction $g_{\text{s}}$ term.
The magnetic field coefficients ($p$ and $q$) and the optical potential
$V({\bf r})$ are set to zero in this section.
The kinetic term, the interaction ($g_n$ and $g_s$) terms, and the chemical 
potential $\mu$ are retained.

%------------------------------------------------------------------
\subsection{Positive $g_s$}

\begin{figure}
  \vspace{-2cm}

  \epsfxsize=8cm \epsfbox{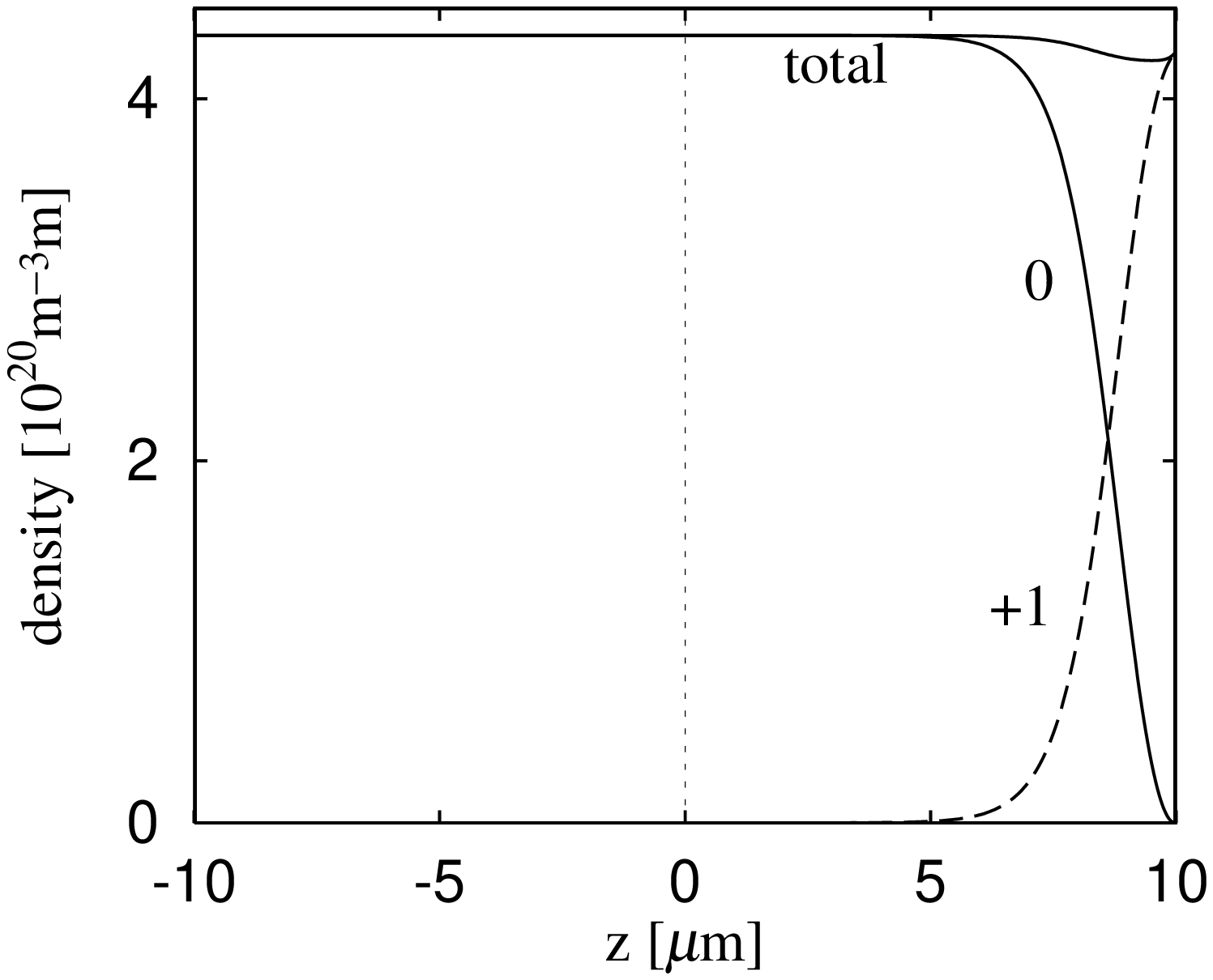} 

  (a)

  \epsfxsize=8cm \epsfbox{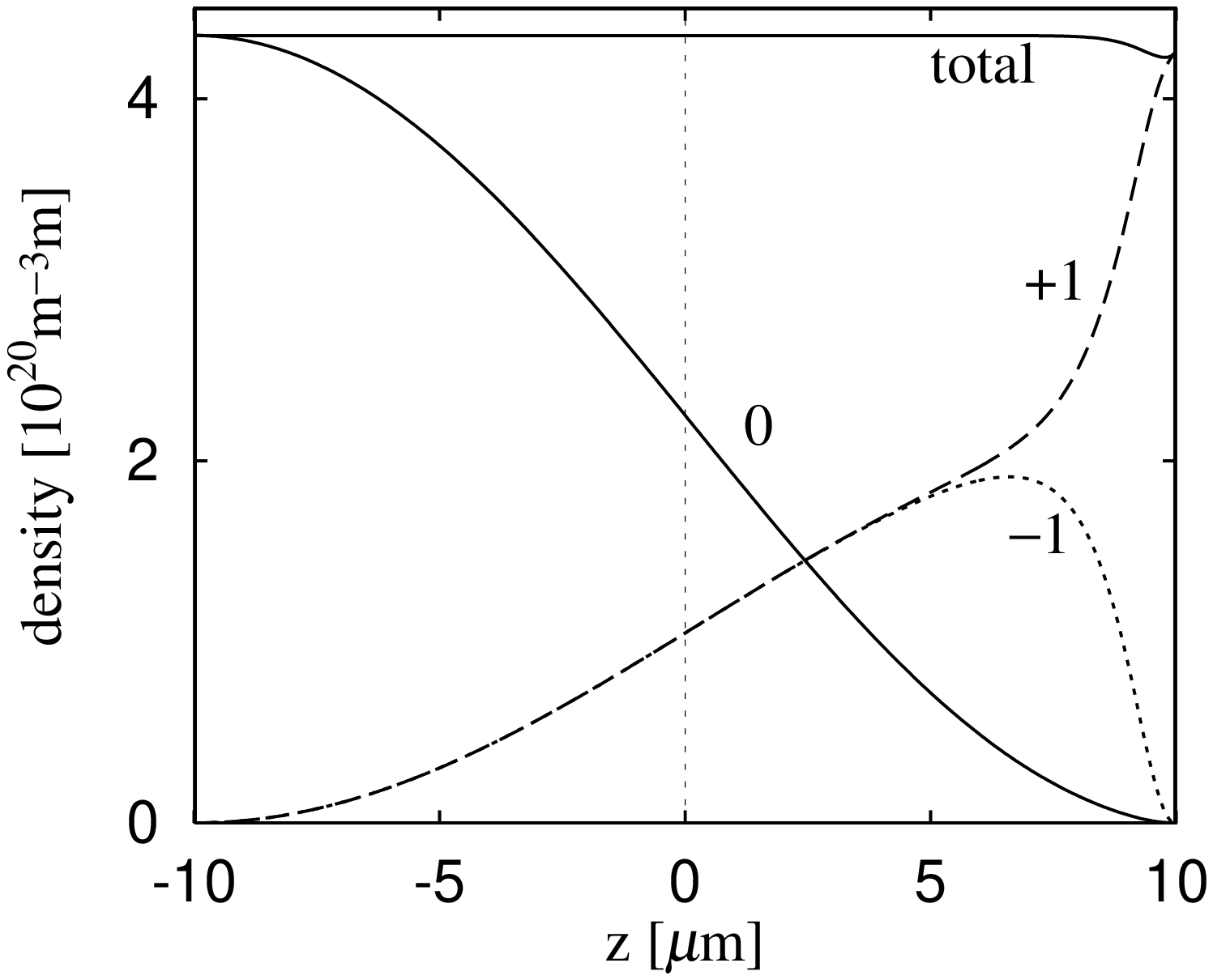} 

  (b) 

  \epsfxsize=7cm \hspace{1cm}\epsfbox{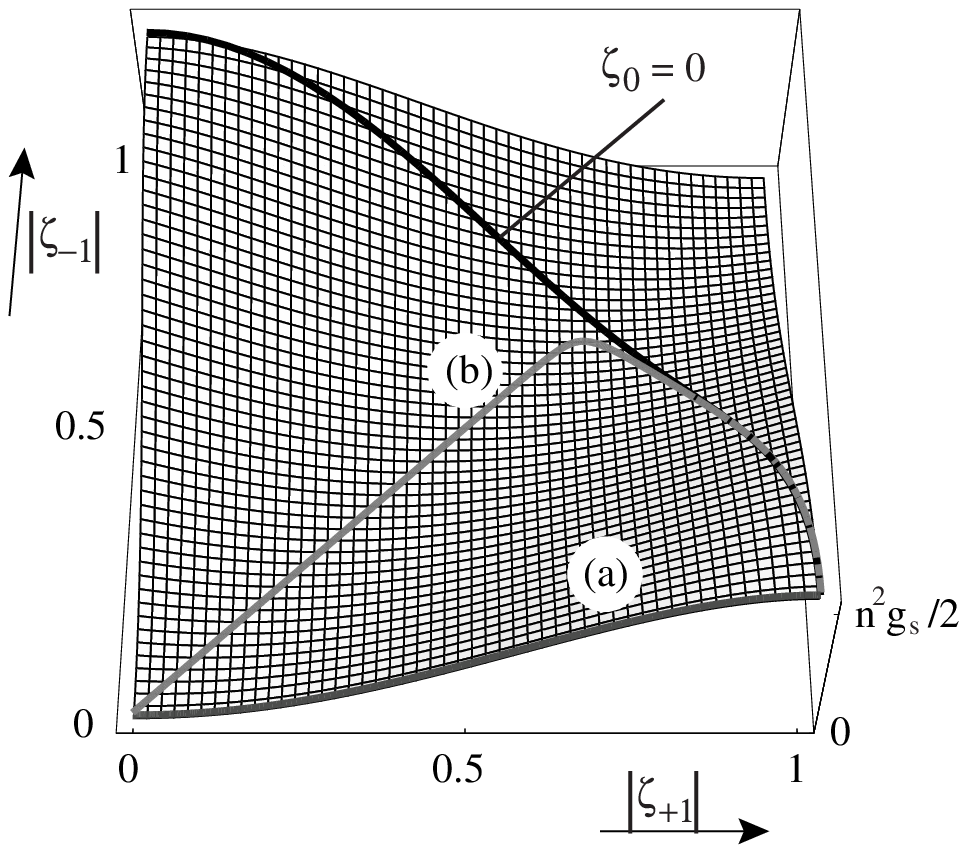} 

  (c)
  \caption{
    The one-dimensional systems with positive
    $g_s$ ($a_0 = 2.75$nm, $a_2 = 2.94$nm).
    The chemical potential $\mu$ is $4.35 \times 10^{20} m^{-3} \times g_n$.
    (a) The density variation of the three components
    when the boundary condition 
    $(|\zeta_{+1}|,|\zeta_{0}|,|\zeta_{-1}|) = (1, 0, 0)$ at $z=-10\mu m$
    and $(0, 1, 0)$ at $z=10\mu m$ is imposed.
    We suppress $\zeta_{-1}$ over whole range of $z$.
    (b) The density variation of the three components
    when the boundary condition is imposed that
    $(|\zeta_{+1}|,|\zeta_{0}|,|\zeta_{-1}|) = (1, 0, 0)$ at $z=-10\mu m$
    and $(0, 1, 0)$ at $z=10\mu m$.
    (c) The landscape of $E_s$.
    The labels (a) and (b) correspond to
    the above figures (a) and (b) respectively.
  \label{fig:1d-lr01}
  }
\end{figure}

Figures\ \ref{fig:1d-lr01}(a) and (b) show the cases when $g_s$ is positive.
The ratio of the components is fixed at each end of the system;
$(\zeta_{+1}, \zeta_{0}, \zeta_{-1})=(0,1,0)$ at the left hand edge
and $(1,0,0)$ at the right hand edge of the system. 
We impose an additional condition that the $\zeta_{-1}$ component
is zero in Fig.\ \ref{fig:1d-lr01} (a).
In this case, the overlapping between $\zeta_{0}$ and  $\zeta_{+1}$ is seen to be an
order of a few micrometers.
This length scale can be explained by 
 $\xi_{s} \equiv \sqrt{\frac{3}{8\pi n (a_2 - a_0)}}$,
 which is about $1.2 \mu m$ in the present case.
Figure\ \ref{fig:1d-lr01} (b) shows that when no component is suppressed,
the overlapping region becomes wider and is subdivided
into the wider region ($-10\mu m<z<5\mu m$) where
$\zeta_{-1} = \zeta_{+1}$ and the narrow region 
($5\mu m < z < 10\mu m$) where $\zeta_{-1} \neq \zeta_{+1}$.
The latter region is characterized by the coherence length $\xi_{s}$. 

The difference between
Figs.\ \ref{fig:1d-lr01}(a) and (b)
is explained by considering the $E_s$ 
term of the total energy [Eq.\ (\ref{eq:Ene})].
The relative phases of $\zeta$'s (equivalently, of $\phi$'s) are determined
under the condition that $E_s$ is minimized as shown in Appendix \ref{sec:phasePhi}.
We can write $E_s$ as 
\begin{eqnarray}
  E_s
&=&
  \frac{n^{2}({\bf r}) g_s}{2} \sum_{\alpha}
    \left(
      \sum_{i,j}
      \zeta_i^{\ast}({\bf r})
      (F_{\alpha})_{i,j}
      \zeta_j({\bf r})
    \right)^2         \nonumber
\\&=&
  \frac{n^2 g_s}{2}
  \bigl[
    -(|\zeta_{+1}| \pm |\zeta_{-1}|)^4 +2 (|\zeta_{+1}| \pm |\zeta_{-1}|)^2
  \bigr].    \label{eq:gsTerm}
\end{eqnarray}
The upper and lower signs correspond to the positive and negative $g_s$ respectively.
Equation\ (\ref{eq:gsTerm}) shows that
%
%------------may11
% the $E_s$ lay the minimum line at
$E_s$ is minimum on the line
%------------may11
%
$|\zeta_{+1}| = |\zeta_{-1}|$ when $g_s > 0$, and
on $|\zeta_{+1}| + |\zeta_{-1}| = 1$ when $g_s < 0$.

Figure\ \ref{fig:1d-lr01}(c) shows the landscape of $E_s$ when $g_s > 0$.
The gray lines (a) and (b) in Fig.\ \ref{fig:1d-lr01}(c) correspond to Figs.\ \ref{fig:1d-lr01}(a) and (b) respectively.
The line (b) goes along the minimum energy line $|\zeta_{+1}| = |\zeta_{-1}|$
at first and
this corresponds to the $z = -10\mu m \text{ to } 5\mu m$ region in
Fig.\ \ref{fig:1d-lr01}(b).
Both the lines (a) and (b) climb up the hill in the $E_s$ landscape from $E_s = 0$ to $E_s = n^2g_s/2$.
This process is controlled by the kinetic term and
the interaction $g_s$ term in the GP equation Eq.\ (\ref{eq:gp}),
and the length scale is an order of $\xi_s$.
% We have an equivalent discussion
% using ``$d$-vector''.

%%%
%%%
%%%
The spatial variation in Fig.\ \ref{fig:1d-lr01}(b) is also understood by
so-called $d$-vector \cite{vollhardt}, which expresses the spin structure of
superfluid.
At $z=-10\mu m$ and $z=10\mu m$ the $d$-vectors are given by
$d=\hat{z}$ and $d = \hat{x} + i \hat{y}$ respectively.
In the region from $z=-10\mu m$ to $5\mu m$ 
the $d$-vector rotates from the $z$-direction to the $x$-direction.
Since these states are energetically degenerate,
this length scale is governed by the system size.
On the other hand, in the region from $z=5\mu m$ to $z=10\mu m$
the $d$-vector is described by
$d = \hat{x} + it\hat{y}$ 
where $t$ varies from $t=0$ $(z=5\mu m)$ to $t = 1$ $(z=10 \mu m)$.
This state change produces the energy variation related
to the interaction $g_s$,
thus is governed by the spin coherence length $\xi_s$.
%%%
%%%
%%%

%------------------------------------------------------------------
\subsection{Negative $g_s$}

\begin{figure}
  \vspace{-2cm}

  \epsfxsize=8cm \epsfbox{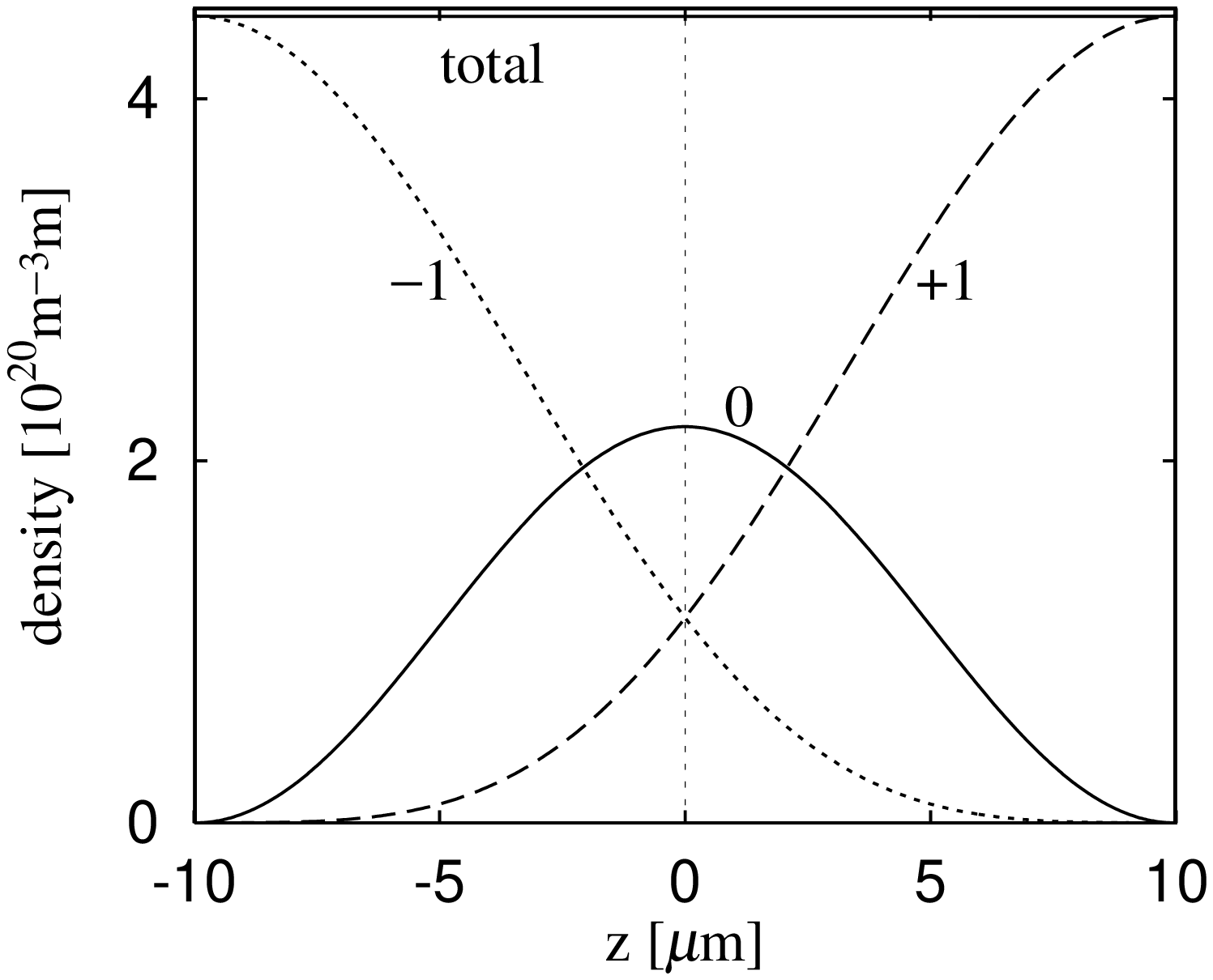} 

  (a) 

  \epsfxsize=8cm \epsfbox{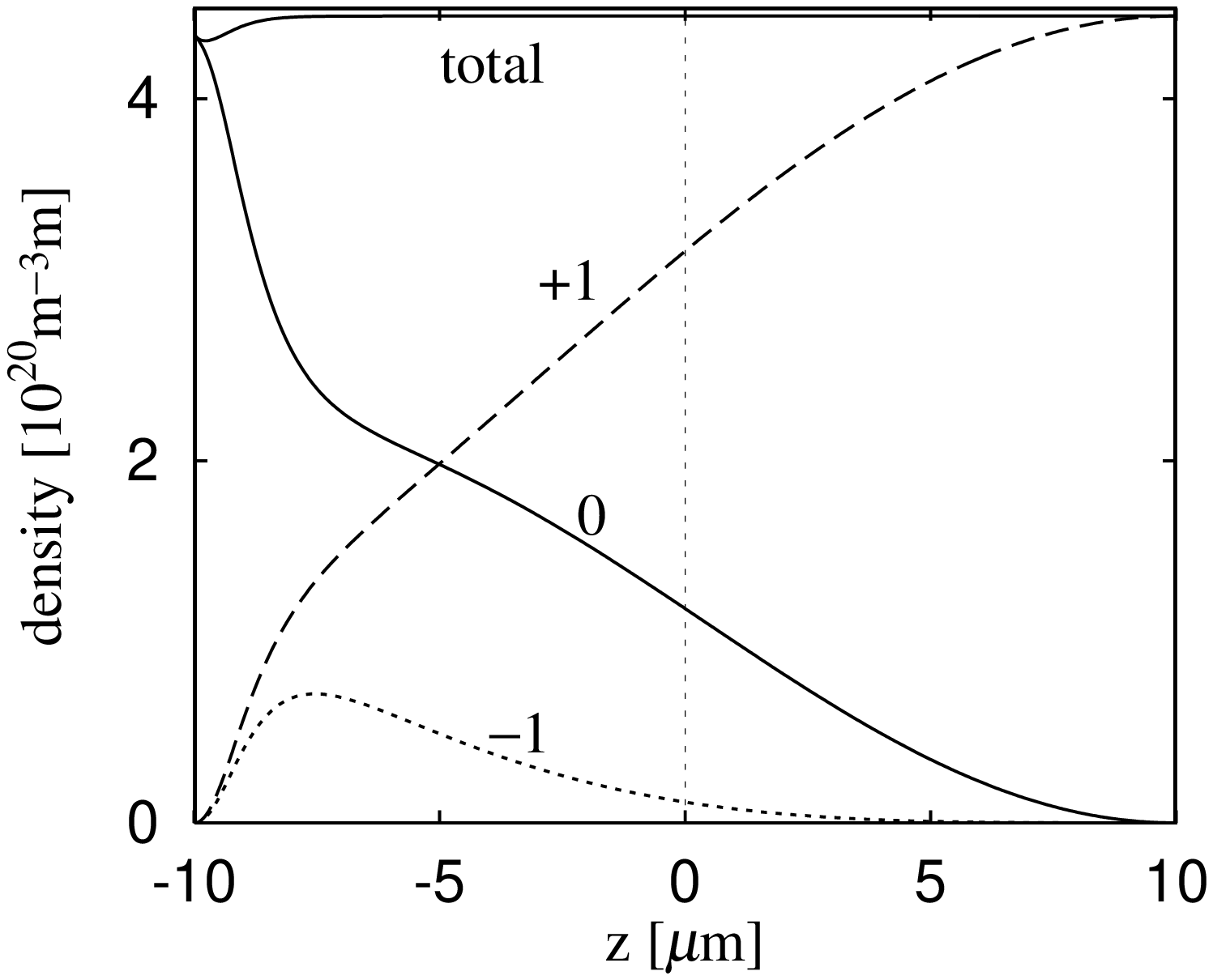} 

  (b) 

  \epsfxsize=7cm \hspace{1cm}\epsfbox{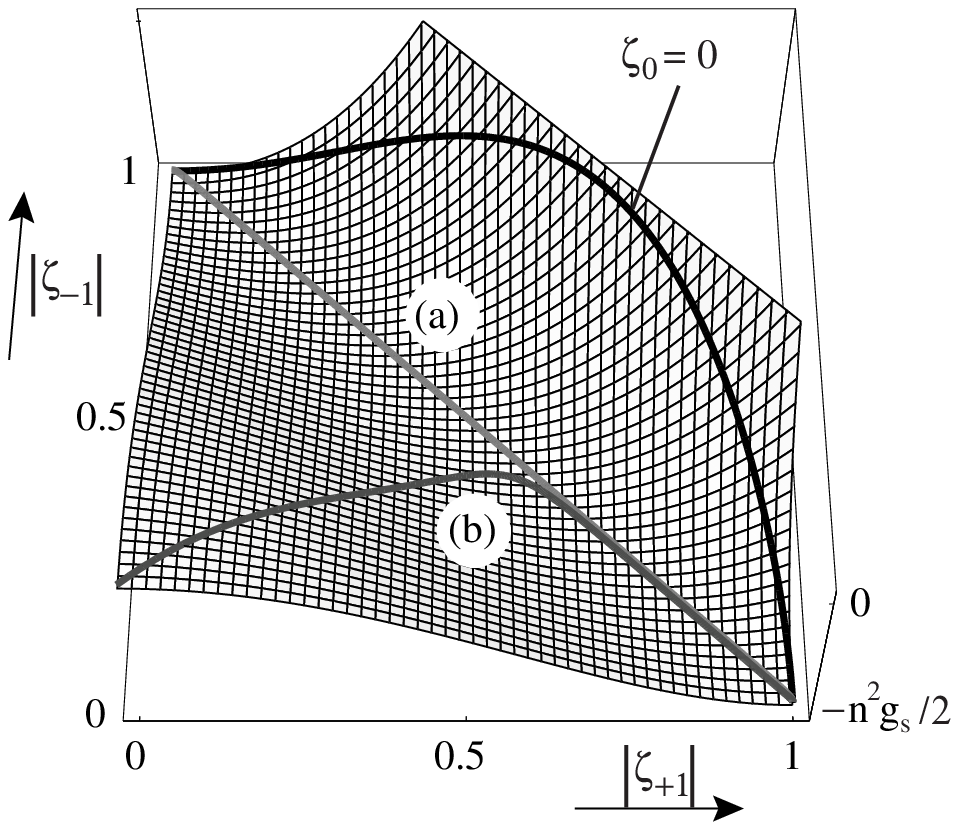} 

  (c) 
  \caption{
    The one-dimensional systems with
    negative $g_s$ $(a_0 = 2.75\text{nm}, a_2 = 2.56\text{nm})$.
    The chemical potential $\mu$ is $4.35 \times 10^{20} m^{-3} \times g_n$.
    (a) The density variation of the three components
    when the boundary condition is imposed that
    $(|\zeta_{+1}|,|\zeta_{0}|,|\zeta_{-1}|) = (0, 0, 1)$ at $z=-10\mu m$
    and $(1, 0, 0)$ at $z=10\mu m$. 
    (b) The density variation of the three components when the
    boundary condition is imposed that
    $(|\zeta_{+1}|,|\zeta_{0}|,|\zeta_{-1}|) = (0, 1, 0)$
    at $z=-10\mu m$ and $(1, 0, 0)$ at $z=10\mu m$.
    (c) The landscape of $E_s$.
    The labels (a) and (b) correspond to the above figures (a) and (b) respectively.
  \label{fig:1dNEG}
  }
\end{figure}

Figure\ \ref{fig:1dNEG} shows the cases when $g_s$ is negative.
In Fig.\ \ref{fig:1dNEG} (a), the ratio of the components is fixed to
$(|\zeta_{+1}|, |\zeta_{0}|, |\zeta_{-1}|)=(0,0,1)$
at the left hand edge and $(1,0,0)$ at the right hand edge of the system. 
As shown by the gray line (a) in Fig.\ \ref{fig:1dNEG} (c), the system goes 
along the minimum $E_s$ line $|\zeta_{+1}| + |\zeta_{-1}| = 1$.
Therefore the length scale is not controlled by $\xi_s$ and it becomes 
as long as the boundary conditions allow, that is, the length scale is determined by the boundary condition.
This gentle shape contrasts remarkably with Fig.\ \ref{fig:1dNEG} (b) which
shows the system when the ratio of the
components is fixed to $(|\zeta_{+1}|, |\zeta_{0}|, |\zeta_{-1}|)=(0,0,1)$ 
at the left hand edge of the system.
The quick change around $z=-10\mu m \text{ to} -7\mu m$ is controlled by $\xi_s$ in a similar 
reason of the previous $g_s>0$ case.
It is interesting to note that the $\zeta_{-1}$ component spontaneously appears to minimize $E_s$;
the system without restriction does not go along the $|\zeta_{-1}| = 0$ line.

%------------------------------------------------------------------
\subsection{Length scales}

The length scale of the density variation in the system {\it without}
the spin freedom is determined by
$\xi=\sqrt{1/(8\pi n a)}$ where $a(>0)$ is the s-wave scattering length.
This is equivalent to
$
\xi_{n} \equiv \sqrt{\frac{3}{8\pi n (2 a_2 + a_0)}}
(2a_2 + a_0 > 0)
$
in the system treated in this paper. 
The origin of $\xi_n$ is the competition
between the kinetic term and the interaction term
with $g_n$ in the GP equation. 

There are three types of the length scale in this system;

(1) $\xi_n$.
This becomes dominant when the total density varies.
%
%-----------
This is not significant in these one-dimensional systems
because their densities are almost uniform.
%-----------
%
In three-dimensional calculations,
the $total$ density profile along the $r$ axis in Fig.\ \ref{fig:3dGPdns},
which is more gentle than that in 3dTF (no corresponding figure in this paper),
is explained by $\xi_n$.

(2) $\xi_s$.
When the total density does not vary significantly, this length scale
becomes important.
The origin is the competition between the $g_s$ term and
the kinetic term in the GP equation. 
For example, Fig.\ \ref{fig:1d-lr01}(a) and the left hand side of Fig.\ \ref{fig:1dNEG}(b).

(3) The length controlled by neither $\xi_n$, nor $\xi_s$.
When the spin state of the system varies along the lowest energy line of $E_s$,
the length scale of the density variation is governed neither by
$\xi_n$, nor  by $\xi_s$.
The characteristic length is controlled by the boundary conditions
in the one-dimensional cases:
The left hand side of Fig.\ \ref{fig:1d-lr01}(b)
and Fig.\ \ref{fig:1dNEG}(a).
They correspond to 
the line (b) in Fig.\ \ref{fig:1d-lr01}(c)
and the line (a) in Fig.\ \ref{fig:1dNEG}(c).
%
%

%--------may11
As for the previous 3-dimensional calculations,
the type (3) of these length scales explains
the difference between the results of the 3dGP and the 3dTF 
in Figs.\ \ref{fig:densGP}(a) and (b).
The magnetic field parameters $p$ and $q$ determine
the characteristic length when 3dGP is used.
It is neither $\xi_n$ nor $\xi_s$.
%
%%  As for the previous 3-dimensional calculations,
%%  the difference between the results of the 3dGP and the 3dTF 
%%  %%% much longer than the $\xi$'s in three-dimensional systems
%%  in Figs.\ \ref{fig:densGP}(a) and (b)
%%  is explained with the type (3) of these length scales.
%--------may11
We note that the $\zeta_{+1}$ and $\zeta_{-1}$ components
have the finite density at $z=0$ in Fig.\ \ref{fig:densGP}(a).
This long length scale is explained by the minimum energy
line of the $g_s$ term in the GP equation (or $E_s$).
This is the $|\zeta_{-1}| = |\zeta_{+1}|$ line of Fig.\ \ref{fig:1d-lr01}(c).

%==================================================================
\section{Summary and Conclusion}

We have investigated the spinor BEC systems based on
the generalized GP equation extended to the cases where
the BEC has the spin degrees of freedom.
The spatial structure of the domain wall which is the interface
of the different spin states of BEC is analyzed by solving the GP equation
both for
rotationally-symmetric (around the $z$ axis)
three-dimensional and one-dimensional cases.
The former case simulates the actual experimental situation of
Stenger {\it et al.}\cite{StengerNature}.
Our calculations show that the simple TF approximation taken by
Stenger {\it et al.} can be justified and yields the correct
value of the interaction constant of the spin channel which turns
out to be antiferromagnetic in the present hyperfine state $F=1$ of $^{23}$Na atoms.
The full GP solutions both in three- and one-dimensional cases reveal
a long length scale associated with the interaction of the spin channel,
yielding the large overlapping region between the immiscible components (e.g.\ 0 and +1).

\section*{acknowledgement}
The authors thank S. Inouye for valuable discussions
on their spin domain experiments.

%===========================================================================
\appendix
\section{Relative Phases of $\phi$}\label{sec:phasePhi}

The relative phases of the three component condensate wavefunctions are
determined so that the energy density of the $g_{s}$ term 
\begin{equation}
  E_{s} =
  \frac{g_s}{2} \sum_{\alpha}
  \Biggl(
    \sum_{i,j} \phi_i^{\ast} (F_{\alpha})_{i,j} \phi_j
  \Biggr)^2    \label{eq:es}
\end{equation}
is minimized under the condition that the amplitude of $\phi$'s are fixed.
This is because the other terms in $E$ are not affected
by the choice of the phases.
We assume that 
\begin{eqnarray}
  (\phi_{ 1}, \phi_{ 0}, \phi_{-1})
  &=&
  (\beta_{ 1} e^{i\gamma_{ 1}}, \beta_{ 0}, \beta_{-1} e^{i\gamma_{-1}})
\end{eqnarray}
where $\beta_{i} (\ge 0)$ and $\gamma_{i}$ are real numbers.
The amplitude $\beta_{i}$ are fixed and we determine $\gamma_{i}$ to minimize $E_s$.
From Eq.\ (\ref{eq:es}) we obtain
%
%
%--------------------------------------------------
\begin{eqnarray}
  \frac{2E_s}{g_s} 
  &=&
  \frac{\phi_{0}^2}{2}[  \left(
    \phi_{1}^{\ast} + \phi_{1} + \phi_{-1} + \phi_{-1}^{\ast} 
  \right)^2            \nonumber
\\&&
  -\left(
    \phi_{1}^{\ast} - \phi_{1} + \phi_{-1} - \phi_{-1}^{\ast}
  \right)^2 ]            \nonumber
\\&&
  + (|\phi_{1}|^2 - |\phi_{-1}^2)|^2      \nonumber
\\
  &=&
  \frac{\beta_{0}^2}{2}[  \left(
    2\beta_{1} \cos \gamma_{1} + 2\beta_{-1} \cos \gamma_{-1}
  \right)^2            \nonumber
\\&&
  +\left(
    2 \beta_{1} \sin\gamma_{1} - 2\beta_{-1} \sin\gamma_{-1}
  \right)^2 ]            \nonumber
\\&&
  + (\beta_{1}^2 - \beta_{-1}^2)^2      \nonumber
\\&=&
  2\beta_{0}^2 [
    \beta_{1}^2 + \beta_{-1}^2
    + 2 \beta_{1} \beta_{-1} \cos(\gamma_{1} + \gamma_{-1})
  ]       \nonumber
\\&&
  + (\beta_{1}^2 - \beta_{-1}^2)^2.
\end{eqnarray}
%--------------------------------------------------
%%%%%%%%%%%%%%%%%%%%%%%%%%%%%%%%%%%%%%%%%%%%%%%%
% \begin{eqnarray}
%   E_s
% &=&
%   \frac{n^2 g_s}{2} \sum_{\alpha} \left(
%       \sum_{i,j} \zeta_i^{\ast}({\bf r}) (F_{\alpha})_{i,j} \zeta_j({\bf r})
%     \right)^2         \nonumber
% \\&=&
%   \frac{n^2 g_s}{2} \bigl[
%     -(|\zeta_{+1}| \pm |\zeta_{-1}|)^4 +2 (|\zeta_{+1}| \pm |\zeta_{-1}|)^2
%   \bigr]    
% \end{eqnarray}
%%%%%%%%%%%%%%%%%%%%%%%%%%%%%%%%%%%%%%%%%%%%%%%%
%
When $g_s > 0$, $E_s$ is minimized with $\gamma_{1} + \gamma_{-1} = \pi$.
Therefore, we can take $\phi$ as
\begin{equation}
  (\phi_{ 1},\phi_{ 0},\phi_{-1}) = (\beta_{1}, \beta_{0}, - \beta_{-1}).
\end{equation}
When $g_s < 0$, $\gamma_{1} + \gamma_{-1} = 0$. Therefore,
\begin{equation}
  (\phi_{ 1},\phi_{ 0},\phi_{-1}) = (\beta_{1}, \beta_{0}, \beta_{-1}).
\end{equation}

%===========================================================================
%%% \section{$\lowercase{d}$-vector}\label{sec:dvector}
%%% 
%%% The spatial variation in Fig.\ \ref{fig:1d-lr01}(b) is understood by
%%% so-called $d$-vector \cite{vollhardt}, which expresses the spin structure of
%%% superfluid.
%%% %
%%% %
%%% At $z=-10\mu m$ and $z=10\mu m$ the $d$-vectors are given by
%%% $d=\hat{z}$ and $d = \hat{x} + i \hat{y}$ respectively.
%%% In the region from $z=-10\mu m$ to $5\mu m$ 
%%% the $d$-vector rotates from the $z$-direction to the $x$-direction.
%%% Since these states are energetically degenerate,
%%% this length scale is governed by the system size.
%%% On the other hand, in the region from $z=5\mu m$ to $z=10\mu m$
%%% the $d$-vector is described by
%%% $d = \hat{x} + it\hat{y}$ 
%%% where $t$ varies from $t=0$ $(z=5\mu m)$ to $t = 1$ $(z=10 \mu m)$.
%%% This state change produces the energy variation related to the interaction $g_s$, thus
%%% is governed by the spin coherence length $\xi_s$.

%===========================================================================

%==================================================================

%==================================================================
\newpage
%------------------------------------------------------------------
%------------------------------------------------------------------
%------------------------------------------------------------------
%------------------------------------------------------------------
%------------------------------------------------------------------
%------------------------------------------------------------------
%------------------------------------------------------------------
\clearpage
\clearpage
%------------------------------------------------------------------

%------------------------------------------------------------------
%------------------------------------------------------------------

\newpage
%------------------------------------------------------------------
%------------------------------------------------------------------

\newpage
%------------------------------------------------------------------
%------------------------------------------------------------------
%==================================================================

\end{document}